# Technological evaluation of two AFIS systems[1]

Marcos Faundez-Zanuy
Escola Universitaria Politècnica de Mataró
Avda. Puig i Cadafalch 101-111
08303 MATARO (BARCELONA) SPAIN
E-mail: faundez@eupmt.es  http:www.eupmt.es/veu

**ABSTRACT**
This paper provides a technological evaluation of two Automatic Fingerprint Identification Systems (AFIS) used in forensic applications. Both of them are installed and working in Spanish police premises. The first one is a Printrak AFIS 2000 system with a database of more than 450,000 fingerprints, while the second one is a NEC AFIS 21 SAID NT-LEXS Release 2.4.4 with a database of more than 15 million fingerprints.
Our experiments reveal that although both systems can manage inkless fingerprints, the latest one offers better experimental results.

**FINGERPRINT RECOGNITION**
According to comparative market shares of different biometric technologies, fingerprint is one of the most matured and the leading one by large. Figure 1 compares the revenues generated by several different biometric technologies [1] and their evolution along time. We can observe the high number of commercial applications that rely on fingerprint authentication.

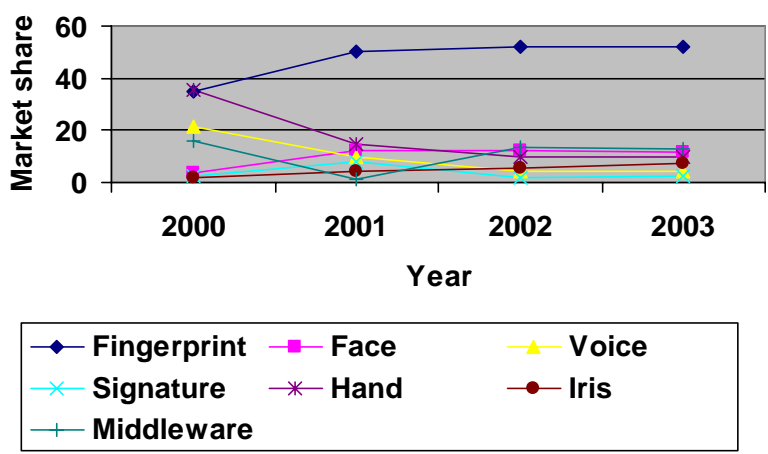

**Fig. 1. Biometric market report (International Biometric Group ©) along several years. These figures exclude AFIS revenues (note that AFIS are used in forensic applications).**

Although police fingerprint applications rely on rolled fingerprint acquisition and civil commercial application use flat fingerprint acquisition, we showed [2] that it is possible to combine both worlds. Mainly, we checked the possibility to match a flat fingerprint acquired with an inkless optical sensor with an ink-rolled fingerprint. In [3] we dealt with an operational report of a civilian fingerprint-based door-opening system described in detail in [4]. We also stated the importance of getting comparative studies from impartial entities without economical interests on a given product. In this paper, we present a technological evaluation of two different Automatic Fingerprint Identification Systems (AFIS), which are fully operative in Spain for forensic applications:
1. Printrak AFIS 2000 system (Printrak international was acquired by Motorola in 2000 [5]) with a database of more than 450,000 fingerprints. It is used by the Catalan state police Policia-Mossos d'Esquadra.
2. NEC AFIS 21 SAID NT-LEXS Release 2.4.4 [6] with a database of more than 15 million fingerprints. It is used by the Spanish Guardia Civil. This same system is shared with the Spanish Dirección General de Policía.

[1] This work has been supported by FEDER and MCYT, TIC-2003-08382-C05-02

As a reference, it is interesting to realize that the AFIS system at FBI consists of a large database of 46 million "ten prints" and conducts, on an average, approximately 50,000 searches per day [7]
One of the goals of this study has been to recommend them about which one seems to be a more powerful supporting tool for forensic scientists in their fight against crime.

This kind of test is intrinsically more difficult to be done than the civilian ones, due to the restricted amount of installed systems, which are much more priced valued, and restricted to civilian researchers without special permission.

**TECHNOLOGY EVALUATION OF TWO DIFFERENT AFIS**
Although there are neither standard rules nor an official homologation laboratory, some best practices in testing and reporting performance of biometric devices exist [8]. According to them, three different testing levels can be established:
1. Technology evaluation
2. Scenario evaluation
3. Operational evaluation

Our study belongs to the first class, where the goal of a technology evaluation is to compare competing algorithms from a single technology. Testing of all algorithms is carried out on a standardized database collected by a "universal" sensor. Nonetheless, performance with this database will depend upon both the environment and the population in which it is collected. Testing is carried out using offline processing of the data. Because the database is fixed, the results of technology tests are repeatable. However, in our case, although the test data from the person used in the experiments is the same, the other fingerprints are different in number and their identity.

**Database and experiments**
We have used the same test database described in [2]. The interested reader can find there a detailed description and photos of the database and Printrak AFIS. The differences between both systems tested enable some tests to be performed just in one system. This is mainly due to the restrictions of the Printrak system, which:
  a) It is unable to successfully extract the characteristic points of the fingerprints (minutiae) in a fully automatic fashion, so they must be manually introduced by a forensic expert. This should not be a surprise. According to [7], many AFIS operations are currently supervised by human experts. FBI can process ~ 16% of the test images in the "lights out" mode – accept AFIS decisions without any manual inspection.
  b) It cannot accept bmp files. Thus, fingerprints must be entered with a scanning procedure using a high resolution camera.

The NEC system, shown in figure 2, does not present these restrictions, so more experiments can be done.

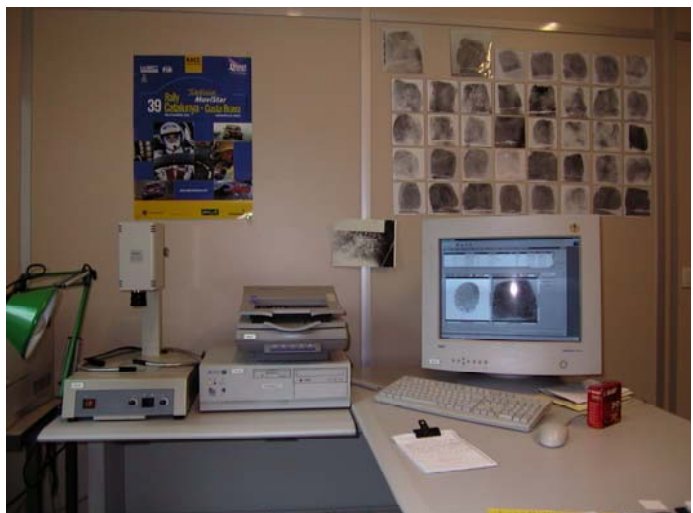

**Fig. 2 NEC AFIS 21 SAID NT-LEXS Release 2.4.4 system**

**RESULTS**

We have worked out two different set of experiments:
a) <u>Technological evaluation of both AFIS:</u> The first experiment consists of the technological evaluation in conditions supported by both AFIS. This implies that the bmp files obtained with the inkless sensor U.are.U [9] are printed on paper and, then they enter the system by means of a scanning procedure. This scanning is based on a video Camera and a scanner (which is a commercial scanner for documents) for the Printrak and NEC AFIS systems respectively. This situation is analogous to those latent fingerprints lifted from crime scenes found on ocular inspections. In these cases, a photo is taken from the fingerprint and enters the AFIS through a scanning procedure.
b) <u>Study of the relevance of the scanning procedure:</u> The second one consists of the comparison between a fully automatic mode, where the bmp file obtained with the inkless sensor U.are.U enters the AFIS, rather than being printed and scanned. This study has only been done with the NEC system. This situation is suitable for remote identification applications, such as those stated in [2].

In all those situations, the machine has been operated by a policeman, which is an expert on fingerprint recognition (each system by its own operator). It is important to take into account that the fingerprints that do not belong to the genuine person used for the test are different in both database (different number and different identities).

**Technological evaluation**

While the Printrak AFIS system was unable to extract the minutiae points and yield a satisfactory result, this has not been a problem for the NEC system. However, according to the forensic scientist, although machines are faster to obtain this set of points, human operators outperform machines, which frequently fail to differentiate ridge endings due to low quality acquisition /fingerprint from real endings. Thus, automatic extraction implied a high set of minutiae, some of them being wrong detections. Taking into account the satisfactory results of the automatic extraction of the NEC AFIS, we have not performed any manual extraction or supervision in this system, which should suppose even better results for the NEC AFIS. Figure 3 shows a snapshot of the NEC AFIS screen.

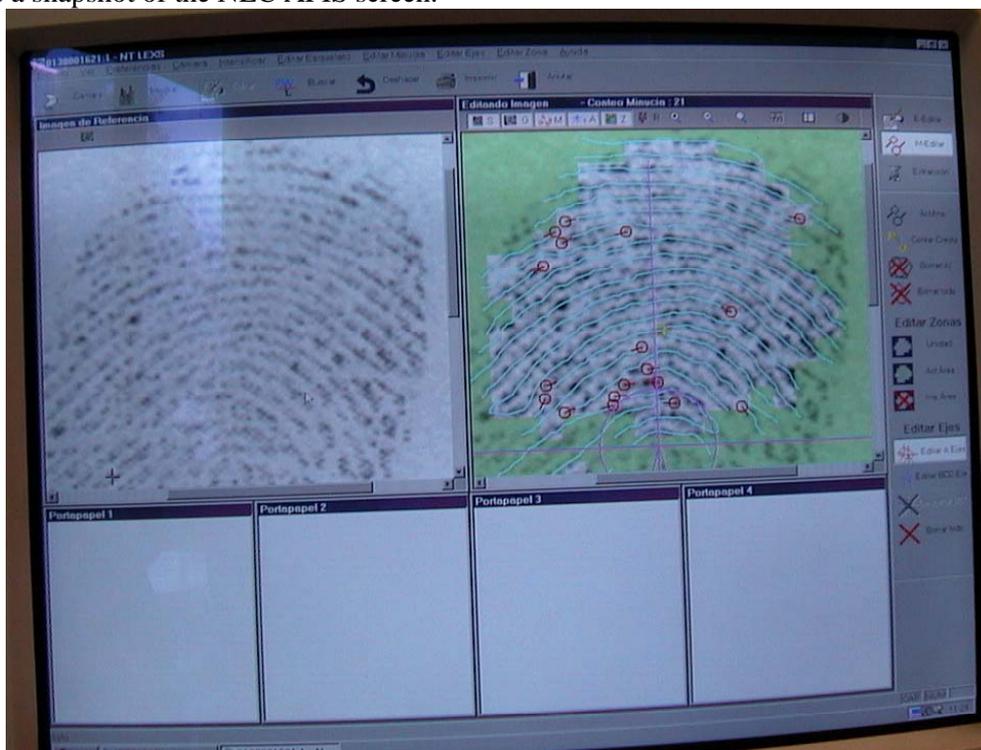

**Fig.3 A screen of the NEC AFIS with the results of the automatic minutiae extraction for trial 1.**

Table 1 compares the scores obtained in each test for the fingerprint, and the score of the next one. The experiment with the Printrak AFIS has been performed with a database of 450,000 fingerprints (we have used the 10 fingers per person contained in the database). The experiment with the NEC AFIS has been performed with a database of 1,500,000 fingerprints (we just used the index finger of each person contained in the database). These databases belong to different individuals, because they have been obtained independently.

**Table 1. Comparison of the scores for the different trials. The input column shows the score obtained by the input trial. For the correct identified fingerprint, the next option means the second candidate. For incorrect ones it means the first candidate.**

|         | Printrak |             |          | NEC   |             |          |
|---------|----------|-------------|----------|-------|-------------|----------|
| Trial # | input    | Next option | Correct? | input | Next option | Correct? |
| 2       | 2160     | 1706        | Yes      | 2988  | 1190        | Yes      |
| 3       | 1625     | 1695        | No       | 2179  | 507         | Yes      |
| 4       | 2705     | 1870        | Yes      | 3408  | 967         | Yes      |
| 5       | 2205     | 1990        | Yes      | 6047  | 785         | Yes      |
| 6       | 2535     | 1495        | Yes      | 5279  | 1248        | Yes      |
| 7       | 2200     | 1675        | Yes      | 4306  | 1406        | Yes      |
| 8       | 1705     | 1765        | No       | 3105  | 542         | Yes      |
| 9       | 1950     | 2090        | No       | 9999  | 1698        | Yes      |
| 0       | 4185     | 2725        | Yes      | 5236  | 1409        | Yes      |

It is interesting to observe that:
- The test number one (trial # 1) does not have enough quality, so it has failed on both systems (the score obtained when comparing with the fingerprint of the same person is not ranked between the 30 highest scores, so it is wrongly assigned to another person). This test is not showed on table 1.
- While the NEC AFIS can correctly assign the identity of the genuine person, the Printrak fails in 3 of 9 tests. However, these results can be conditioned by the skill of the human operator for the Printrak AFIS, at least in some trials.
- The scores obtained with Printrak are lower than and closer to the next candidate, so the result of the NEC system offers more reliability about whom the input fingerprint belongs to.
- Although all the fingerprints used for testing look similar (see fig. 3 of AES [2]) it seems clear that there are certainly differences on the scores obtained, especially for the fully automatic operation mode of NEC AFIS. Thus, the correlation between consecutive acquisitions of a same finger is not as high as common sense says. Consequently, it seems interesting for fingerprint applications to perform several acquisitions rather than rely on a single one.

**Study of the relevance of the scanning procedure, human supervision and the database size**

Another set of experiments has consisted of the evaluation of the relevance of the database size, the scanning procedure, and the intervention of a human supervisor. We have used the bmp files provided by the U.are.U optical sensor, and tried to match them to the whole database (1,500,000 people) using 10 fingers per person (15,000,000 comparisons) and just the index finger (1,500,000 comparisons). We also studied a fully automatic method without human supervision, and with set of features manually extracted by a forensic scientist.

The main conclusions of this experiment are:
- There is a slightly better performance when directly using the bmp files (compare tables 1 and 2). Even the trial number 1 that failed with the scanning procedure is close to be solved (the genuine identity is ranked on the fifth position, while the scanning procedure did not include it among the first 30 positions).
- The database size has relevance on the scores of the next candidates, but in our case, it has not reduced the ability of the system to identify the correct person.

- Comparing the last columns of table 2 it is clear that a human operator outperforms the automatic mode. Thus, the state of the art systems are a powerful tool to help human beings, but they cannot replace us. The necessity of an intervention of a human operator is expected to last for a long time.

**Table 2. Comparison of the scores for the different trials, using the bmp file acquired with the scanner as input. The input column shows the score obtained by the input trial. Next option means the second candidate, except for the first trial.**

| Trial # | NEC 15,000,000 | | NEC 1,500,000 (fully automatic) | | NEC 1,500,000 (manual supervision) | |
|---|---|---|---|---|---|---|
| | input | Next option | input | Next option | input | Next option |
| 1 | | | 316 | 552 | 3213 | 1116 |
| 2 | 1479 | 1311 | 1479 | 819 | 9999 | 1708 |
| 3 | 3854 | 842 | 3854 | 484 | 9999 | 1944 |
| 4 | 3683 | 857 | 3683 | 601 | 9999 | 2278 |
| 5 | 6407 | 1778 | 6407 | 639 | 9999 | 2160 |
| 6 | 7800 | 1918 | 7800 | 762 | 9999 | 1819 |
| 7 | 3360 | 653 | 3360 | 618 | 9999 | 1703 |
| 8 | 6257 | 2879 | 6257 | 1673 | 9999 | 1770 |
| 9 | 7756 | 1819 | 7756 | 1819 | 9999 | 2486 |
| 0 | 5130 | 961 | 5130 | 961 | 9999 | 2537 |

**Acknowledgements**
I want to acknowledge Santi G. Tugores, Jordi Costa, Gabriel Costa-Tarrida and Xavier G. de Linares from the Catalan Policia-Mossos d'Esquadra, and the criminalistic team from the Spanish Guardia Civil for their support and collaboration at the police premises. Indeed, this project would not be possible without their contribution.

**CONCLUSIONS**

Taking into account the results of our experiments, we think that the NEC system offers better identification success and more flexibility for entering fingerprints from the real world, at least, for the tested versions of both systems. The state police were thinking of updating their system, so our recommendation is to adopt the NEC system. This same system has been adopted in 37 central sites in North America and other countries worldwide [10].
Our experiments confirm, in a more scientific way, the personal feeling of the forensic scientist from the Spanish police.